\documentclass[a4paper,jcp,reprint,showpacs,superscriptaddress,amsmath,amssymb]{revtex4-1}

\usepackage{graphicx,color}

\begin{document}

\title{Flexible confinement leads to multiple relaxation regimes in glassy colloidal liquids}

\author{Ian Williams}
\affiliation{H.H. Wills Physics Laboratory, Tyndall Ave., Bristol, BS8 1TL, UK}
\affiliation{School of Chemistry, Cantock's Close, University of Bristol, BS8 1TS, UK}
\affiliation{Centre for Nanoscience and Quantum Information, Tyndall Avenue, Bristol BS8 1FD, UK}
\author{Erdal C. O\u{g}uz}
\affiliation{Department of Chemistry, Princeton University, Princeton, New Jersey, 08544, USA}
\affiliation{Institut f\"ur Theoretische Physik II, Heinrich-Heine-Universit\"at, D-40225 D\"usseldorf, Germany}
\author{Paul Bartlett}
\affiliation{School of Chemistry, Cantock's Close, University of Bristol, BS8 1TS, UK}
\author{Hartmut L\"owen}
\affiliation{Institut f\"ur Theoretische Physik II, Heinrich-Heine-Universit\"at, D-40225 D\"usseldorf, Germany}
\author{C. Patrick Royall}
\email{paddy.royall@bristol.ac.uk}
\affiliation{H.H. Wills Physics Laboratory, Tyndall Ave., Bristol, BS8 1TL, UK}
\affiliation{School of Chemistry, Cantock's Close, University of Bristol, BS8 1TS, UK}
\affiliation{Centre for Nanoscience and Quantum Information, Tyndall Avenue, Bristol BS8 1FD, UK}

\date{\today}

\begin{abstract}
Understanding relaxation of supercooled fluids is a major challenge and confining such systems can lead to bewildering behaviour. Here we exploit an optically confined colloidal model system in which we use reduced pressure as a control parameter. The dynamics of the system are ``Arrhenius'' at low and moderate pressure, but at higher pressures relaxation is faster than expected. We associate this faster relaxation with a decrease in density adjacent to the confining boundary due to local ordering in the system enabled by the flexible wall.
\end{abstract}

\maketitle

\section{Introduction}

The canonical visualisation of dynamical slow-down in supercooled fluids approaching the glass transition is the Angell plot \cite{angell1995}. Approaching the glass transition, the structural relaxation time, $\tau_\alpha$, increases by many orders of magnitude over only a small change in temperature \cite{cavagna2009,berthier2011}. The Angell plot represents this phenomenon, showing the logarithm of $\tau_\alpha$ as a function of inverse temperature. Structural relaxation is related to an energy scale $E$ as $\tau_\alpha \sim \exp (E/k_BT)$ where $k_BT$ is the thermal energy. Constant $E(T)$ corresponds to a straight line on the Angell plot \textit{i.e.} Arrhenius or strong behaviour. Many glassformers exhibit super-Arrhenius or fragile behaviour which corresponds to $E(T)$ increasing upon supercooling, implying more cooperative rearrangements. Recently, a number of materials have been observed to exhibit a change in the dynamic behaviour from fragile to strong at a certain degree of supercooling, a so-called fragile to strong transition in which the low-temperature liquid exhibits strong behaviour \cite{tanaka2005a,saikavoivod2005,mallamace2010}. 

Hard spheres are often employed as a model glass forming system and are readily approximated experimentally using colloids \cite{royall2013myth} and granular matter \cite{watanabe2008,candelier2010,watanabe2011}. These materials can be observed in real-space giving insight into the glass transition at the single-particle level \cite{ivlev,hunter2012,yunker2014}. Such data can help resolve questions about whether the slow dynamics are related to a change in structure (amorphous order) due to the glass transition \cite{royall2008,mazoyer2011,lynch2008,yunker2009,watanabe2008,leocmach2012}. Despite a wealth of results obtained in recent years, whether local structure is a cause or byproduct of the slow dynamics associated with the glass transition remains unresolved \cite{royall2014physrep}.

Hard sphere behaviour depends upon the packing fraction, $\phi_\mathrm{3d}$, and increasing density leads to dynamical arrest \cite{cipeletti2005,brambilla2009} and may be mapped directly to quenching in some systems \cite{gnan2010}. In an analogy between hard spheres and thermal systems, the reduced pressure $Z(\phi) = P/(\rho k_{\mathrm{B}}T)$ where $P$ is pressure and $\rho$ is the number density can be thought of as an inverse temperature \cite{berthier2009,xu2009}. Then divergence of $\tau_\alpha$ at random close packing, where $Z$ diverges, is equivalent to that at $T=0$, as suggested by theories such as dynamic facilitation \cite{chandler2010}. On the other hand, divergence at finite $Z$ implies a glass transition distinct from jamming, corresponding to finite $T$. The consequences for the fragility are noteworthy. Upon changing $\phi_\mathrm{3d}$ from 0.50 to 0.58, $\tau_\alpha$ changes by several decades. Over the same range, $Z$ doubles. As a consequence, when hard sphere relaxation times are plotted as a function of $\phi_\mathrm{3d}$ they can appear to be extremely fragile, while as a function of $Z$ their fragility is comparable to Lennard-Jones models \cite{berthier2009,royall2014}. 

The behaviour of glassforming materials confined to a small pore is strongly dependent upon the pore geometry and the interactions with the pore surface \cite{alcoutlabi2005,richert2011,cammarota2013}. However a coherent picture of the glass transition in confinement is yet to be developed \cite{albasimionesco2006}. Experiments and simulations of molecular, colloidal and model systems report that dynamics in confinement are enhanced \cite{barut1998,arndt1997,pissis1998,morineau2002}, suppressed \cite{morineau2002,nugent2007,scheidler2007,schuller1994,watanabe2011} or unaltered when compared to the bulk, resulting in a decreased, increased or unchanged glass transition temperature (or density) \cite{albasimionesco2006}. Furthermore, local structure can be influenced by the confinement \cite{skinner2013,hunter2014} with significant implications for the dynamics \cite{watanabe2011}. 

Among the most significant impacts of confinement in studies of dynamical arrest is that it can be used to suppress crystallisation. Under confinement, systems such as water, which would otherwise rapidly crystallise, can be vitrified \cite{roussenova2014}. Thus confinement provides a rare experimental route to investigate the controversy over whether water has a liquid-liquid transition (LLT) \cite{mishima1998,mishima2002}, which is among the most hotly contested topics in chemical physics \cite{limmer2011,palmer2014,chandler2014}. Evidence for such LLTs has been obtained in a variety of glassformers \cite{mcmillan2007}. Although LLTs imply a change in local structure, this is often hard to access in experiments on atomic and molecular systems. Some evidence has been obtained in computer simulation \cite{elenius2010,speck2012,speck2014}, but the dynamics related to such transitions are often so slow that special biasing techniques are required \cite{speck2012,speck2014}. In rare examples where unbiased simulations can access the LLT, it has been found to coincide with a fragile-to-strong transition \cite{elenius2010}.

Here we consider the dynamic response of a quasi hard disc system under flexible confinement in both colloidal experiment and computer simulation. In hard discs, the locally favoured structure which becomes more prevalent upon supercooling is hexagonal \cite{kawasaki2007}. We show that considering the dynamical behaviour in terms of reduced pressure rather than area fraction reveals multiple relaxation regimes. Furthermore we identify the structural origin of the observed dynamics and relate it to a transition to a state rich in the hexagonal order of the deeply supercooled liquid. The results we present serve to underline the profound effects the details of a wall can have on the dynamical behaviour a confined material.

\section{Flexible Confinement}

\begin{figure}
\begin{center}
\centerline{\includegraphics[width=80mm,height=63mm]{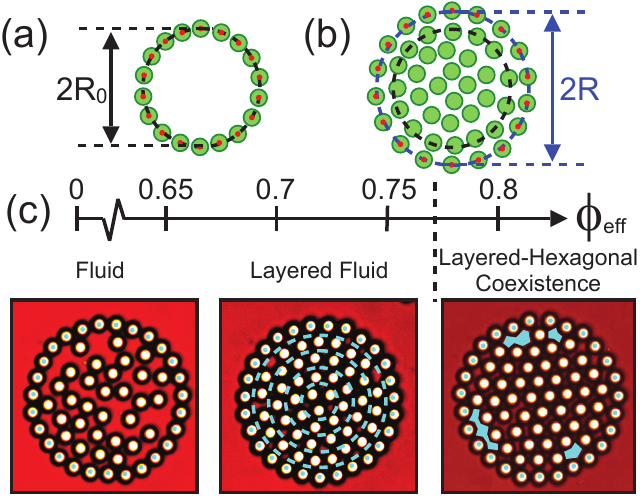}}
\caption{\label{figSchematic} (a) Schematic showing colloidal corral geometry. Optically trapped particles (marked with dots) form a circular boundary of radius $R_0$. (b) Particles confined to the interior stretch the boundary to radius $R > R_0$ resulting in optical restoring forces. (c) Corral state diagram and micrographs illustrating observed structures. Dashed lines denote layers in layered fluid. Large voids in hexagonal configuration are highlighted.}
\end{center}
\end{figure}

Our experimental system consists of polystyrene spheres of diameter $\sigma = 5 \; \mathrm{\mu m}$ and polydispersity $s=0.02$ suspended in a mixture of water and ethanol at a ratio of $3$ to $1$ by weight. Screened electrostatic interactions between these colloids are described by a Debye length of $\lambda_{\mathrm{D}} \approx 25 \; \mathrm{nm}$ and are accounted for by employing a Barker-Henderson effective hard sphere diameter of $\sigma_{\mathrm{eff}} = 5.08 \pm 0.016 \; \mathrm{\mu m}$ \cite{barker1976}. A mismatch in density between the particles and solvent results a gravitational length of $l_g / \sigma_{\mathrm{eff}} = 0.015(1)$  and thus particle sedimentation. For sufficiently dilute samples, a quasi-two-dimensional monolayer is formed adjacent to a glass substrate, exhibiting negligible out-of-plane motion. The glass substrate is made hydrophobic by treatment with Gelest Glassclad 18 to inhibit the adhesion of particles. This experimental system is observed to behave in a hard-disc-like manner \cite{williams2013,williams2014}.

Holographic optical tweezers built around a liquid crystal spatial light modulator and standard inverted microscope are used to create $27$ independently steerable optical traps which are placed on a circle or ``corral'' as shown in Fig. \ref{figSchematic} (a). This ring of particles creates a deformable, circular boundary for additional colloids confined to its interior [Fig. \ref{figSchematic} (b)]. The apparatus is controlled using LabVIEW software developed and distributed by the Glasgow University Optics Group \cite{preece2009}. Upon populating the interior the boundary expands --- the optically trapped wall particles are forced away from their potential energy minima and experience Hookean restoring forces with a spring constant $\kappa = 302(2) \; k_{\mathrm{B}} T \sigma_{\mathrm{eff}}^{-2}$. We determine the spring constant of the flexible boundary by measuring the probability distribution of radial positions of optically trapped particles in the absence of a confined population. Using the Boltzmann relation we thus extract the optical potential and fit this with the parabolic form of a Hookean spring. We find good uniformity in the strength of our optical traps and thus characterise all optical traps with a single spring constant. The osmotic pressure then reads $p = 27 \kappa \frac{(R - R_0)}{2 \pi R}$ where $R$ and $R_0$ are the radii of the populated and unpopulated corrals.

Complementary Monte Carlo simulations of a similarly confined system of hard discs are performed by locating $27$ discs in parabolic potential wells representing the optical traps employed experimentally. Arranging the potential minima on a circle of radius $R_0 = 4.32 \sigma$ creates a confining boundary matching the experimental colloidal corral. Each simulation runs for $10^7$ Monte Carlo sweeps and is sampled every $10^3$ sweeps where a sweep corresponds to an attempted move of each particle including those located in the parabolic traps. The step size for attempted moves is adjusted such that the acceptance rate is approximately $40 \%$.

We exploit our measurements of pressure to consider the response of the relaxation time to the reduced pressure and focus on the effect of the confinement. Thus, in addition to the confinement of the colloidal corral, we simulate a bulk binary system with a size ratio of $1:1.4$ \cite{dunleavy2012}. This system is a $50:50$ mixture of $N=20000$ hard discs of size ratio $1.4$ in a square box with periodic boundaries. This system is evolved via Monte Carlo dynamics in which a trial move involves displacing a random particle to a random position within a window of size $0.05 \times 0.05 \sigma^2$ centred on its original position where $\sigma$ is the small particle diameter. Trial moves are accepted as long as they do not lead to particle overlaps. In these bulk simulations, pressure is extracted from the radial distribution functions, $g_{\alpha \beta}(r)$, using the relationship \cite{thorneywork2014}:
\begin{equation}
\frac{p}{\rho k_{\mathrm{B}}T} = 1 + \frac{\pi}{2} \sum_{\alpha,\beta=1}^2 x_\alpha x_\beta (\sigma_{\alpha \beta} \rho^{1/2})^2 g_{\alpha \beta}(\sigma_{\alpha \beta}^{+})
\end{equation}
where $\alpha$ and $\beta$ label particle species, $\rho$ is the total number density, $x_\alpha = N_\alpha/N$ is the fraction of particle species $\alpha$, $\sigma_{\alpha \beta} = \frac{1}{2}(\sigma_\alpha + \sigma_\beta)$ is the pairwise additive hard core diameter and $g_{\alpha \beta}(\sigma_{\alpha \beta}^{+})$ is the value of the radial distribution function at $r = \sigma_{\alpha \beta}$. Further Monte Carlo simulations are performed of $N=2500$ monodisperse hard discs in order to compare the degree of hexagonal ordering in the confined system to that in the crystalline bulk.

We show that considering the relaxation behaviour of a colloidal system in terms of reduced pressure rather than area fraction can reveal intriguing new phenomena. In our confined system, the interplay between structure and dynamics results in multiple relaxation regimes. This behaviour is a direct consequence of the deformable nature of our confining boundary. 

\section{Results}

The structural behaviour of the colloidal corral is summarised in Fig. \ref{figSchematic} (c). At low effective area fractions the confined system is fluid-like. On increasing the interior population, configurations consisting of four concentric particle layers develop, reminiscent of the structure of similar systems confined by hard boundaries \cite{nemeth1998}. However, for effective area fractions $\phi_{\mathrm{eff}} \gtrsim 0.77$ corresponding to interior populations $N \ge 47$, the system exhibits a structural bistability between the concentrically layered structure and configurations with enhanced hexagonal ordering. In order to adopt this structure the boundary must undergo deformations away from its circular shape and therefore the observed structural bistability is a consequence of boundary flexibility \cite{williams2014}. Examples of these two configurations are shown in Fig. \ref{figSchematic} (c). For such a small system crystallisation is irrelevant, and as we have noted above many glassforming liquids in 2d, even those with polydispersity $> 9\%$, exhibit regions of enhanced hexagonal order upon supercooling \cite{kawasaki2007,watanabe2008,watanabe2011}. Furthermore, the hexagonal structures are incommensurate with the curved boundary which has significant dynamical implications as described below.

\begin{figure}[htb]
\begin{center}
\centerline{\includegraphics[width=75mm,height=105mm]{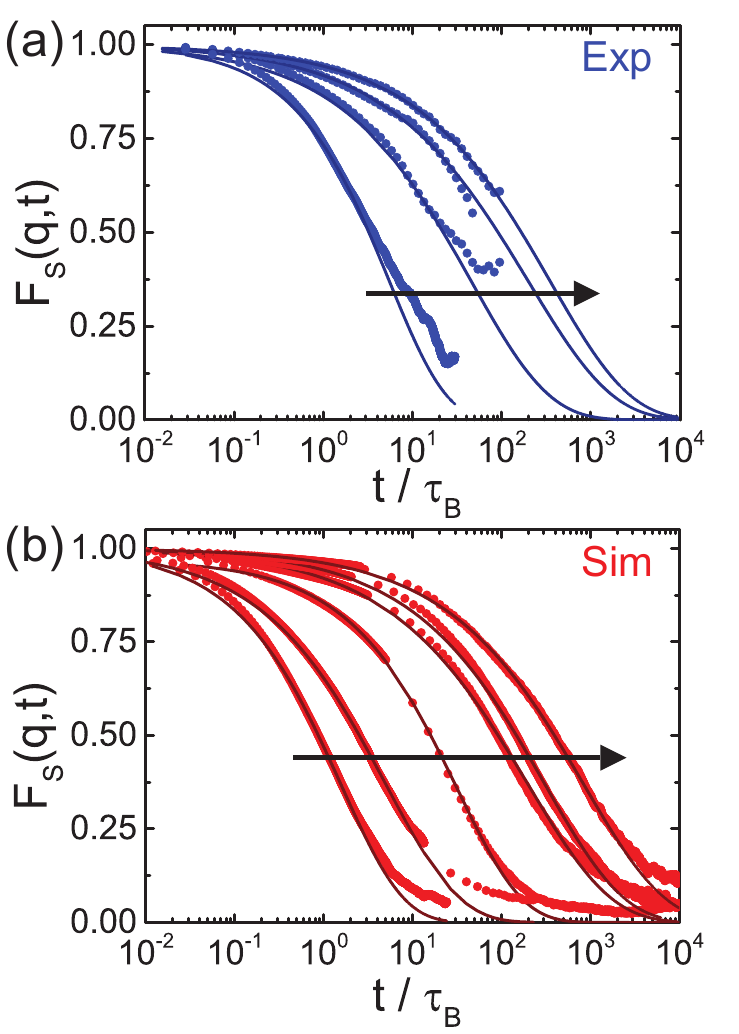}}
\caption{\label{figISFs} (a) Examples of experimentally measured self intermediate scattering functions for populations $N=30$, $N=44$, $N=47$ and $N=49$ (points) and stretched exponential fits (lines). (b) Examples of simulated self intermediate scattering functions for populations $N=30$, $N=37$, $N=44$, $N=47$, $N=49$ and $N=51$ (points) and stretched exponential fits (lines). In both plots, population increases from left to right in the direction of the arrow.}
\end{center}
\end{figure}

The degree of hexagonal ordering is quantified using the bond orientational order parameter $\psi_6$, defined for particle $j$ as $\psi_{6}^{j}  = \left| \frac{1}{z_j} \sum_{m=1}^{z_j} \exp{(i 6 \theta_m^j)} \right|$ where $z_j$ is the co-ordination number of particle $j$, $m$ labels its neighbours and $\theta_m^j$ is the angle made between a reference axis and the bond joining particles $j$ and $m$. The vertical bars represent the magnitude of the complex exponential. Spatial and temporal averaging yields $\psi_6 = \langle \psi_6^j \rangle$. Adjacent to the curved wall, $\psi_6^j$ is always strongly suppressed to a value of $\psi_6^j \approx 0.5$. When characterising structure within the corral with an average $\psi_6^j$ we consider only contributions from particles that are non-adjacent to the boundary such that this wall-curvature-defined value does not dominate the averaging. As such we are able to distinguish locally hexagonal from layered structures using $\psi_6$.

\subsection*{Dynamical Behaviour}

\begin{figure*}[htb]
\begin{center}
\centerline{\includegraphics[width=160mm,height=63mm]{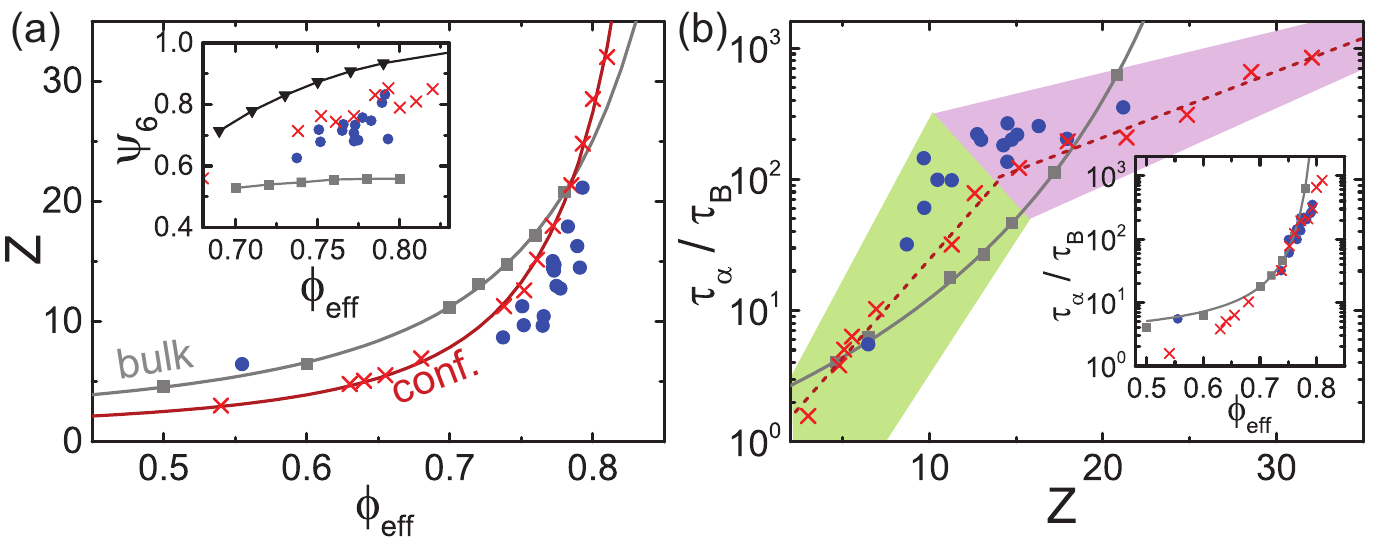}}
\caption{\label{figEOSAngell}  (a) Reduced pressure as a function of effective area fraction as measured in the corral system in experiment (blue points) and simulation (red crosses). The inset shows $\psi_6$ averaged over particles non-adjacent to the boundary as a function of $\phi_{\mathrm{eff}}$ compared to binary hard disc simulations (gray squares) and monodisperse hard disc simulations (black triangles). (b) Relaxation time, $\tau_\alpha$, scaled by Brownian time as a function of reduced pressure measured in experiment (blue points) and simulation (red crosses). Green shaded region represents fluid and layered fluid regime, purple shaded region represents structurally bistable systems. Dashed lines are straight line fits. The inset shows $\tau_\alpha$ as a function of $\phi_{\mathrm{eff}}$. Grey points in both plots are from bulk simulations of a binary hard disc glassformer \cite{dunleavy2012} and grey lines in (b) are VFT fits to these data.}
\end{center}
\end{figure*}

Particle dynamics are characterised through the self part of the intermediate scattering function (ISF) $F_S(q,t) = \langle | \exp \left(i q[r(t+t') - r(t')] \right)| \rangle$ where the angle brackets indicate averaging over all confined particles and $t'$ and the wavevector $q = 2 \pi \sigma_{\mathrm{eff}}^{-1}$. In an ergodic system the ISF fully decays to zero, while in a glass it does not decay on the experimental or simulated timescale. Full ISF decay is observed in our Monte Carlo simulations, and thus our system is considered to be a supercooled fluid, rather than a glass. The structural relaxation time, $\tau_\alpha$, is extracted from a stretched exponential fit to the ISF. Figure \ref{figISFs} shows experimental (a) and simulated (b) intermediate scattering functions (points) and the fits to these data (lines). The relevant timescale for both experiments and simulations is the Brownian time, $\tau_{\mathrm{B}}$, the time taken for an isolated particle to diffuse its diameter. Here $\tau_{\mathrm{B}} = 70.2 \; \mathrm{s}$. Comparison between experimental timescales (in seconds) and simulated timescales (in sweeps) is made using $\tau_{\mathrm{B}}$.

Our experimental set-up enables direct measurement of the equation of state which is shown in Figure \ref{figEOSAngell} (a) as a function of effective area fraction. Also plotted are data from simulations of binary bulk hard discs. Figure \ref{figEOSAngell} (b) shows the relaxation times, $\tau_\alpha$, measured in experiment and simulation as a function of reduced pressure -- the Angell plot. The inset shows the same data plotted as a function of $\phi_{\mathrm{eff}}$. The increase in relaxation time with reduced pressure does not follow the usual behaviour observed for glassformers in which $\log (\tau_\alpha)$ increases linearly or super-linearly upon supercooling for strong and fragile glassformers respectively. Instead, two regimes are evident. Both experimental (blue circles) and simulated data (red crosses) exhibit a kink in the Angell plot at reduced pressure $Z \approx 13$. For $Z \lesssim 13$, relaxation time scales with $Z$ in an Arrhenius-like manner. For $Z \gtrsim 13$, the system relaxes faster than expected from extrapolation of the line implied by the low $Z$ data --- the Angell plot appears to curve \emph{downwards}. In other words, this is the analogue in hard systems where pressure drives arrest, of a fragile-to-strong transition noted in certain atomic and molecular glassformers \cite{tanaka2005a,saikavoivod2005,elenius2010,mallamace2010}.

This change in relaxation behaviour occurs at the onset of structural bistability in the confined system. The ``Arrhenius-like'' scaling of relaxation time at low pressure occurs for fluid-like and purely layered samples at $\phi_{\mathrm{eff}} < 0.77$, while in the bistable regime relaxation occurs more quickly than expected. In other words, the change in relaxation behaviour occurs with the development of local hexagonal ordering. Interestingly, this change in behaviour at $\phi_{\mathrm{eff}} \approx 0.77$ is not immediately obvious when $\tau_\alpha$ is plotted as a function of $\phi_{\mathrm{eff}}$ [Fig. \ref{figEOSAngell} (b) inset]. Clearly, considering the dynamics as a function of reduced pressure rather than density yields a deeper and more sensitive insight into relaxation processes. Here it is important to show that the confined system has a much weaker degree of local hexagonal order than does a bulk system of monodisperse hard discs, which forms an hexagonal crystal for $\phi \ge 0.72$ \cite{bernard2011}. The inset to Fig. \ref{figEOSAngell} (a) shows $\psi_6$ averaged over all particles that are non-adjacent to the boundary as measured in corral experiment (blue circles) and simulation (red crosses) as a function of effective area fraction. By considering only particles in the inner region, average $\psi_6$ is enhanced as the boundary strongly suppresses hexagonal ordering in its vicinity. These data are compared to bulk Monte Carlo simulations of a binary hard disc glassformer (gray squares) and monodisperse hard discs, which form an hexagonal crystal (black triangles).

\subsection*{Mechanism of relaxation}

\begin{figure*}[htb]
\begin{center}
\centerline{\includegraphics[width=160mm,height=52mm]{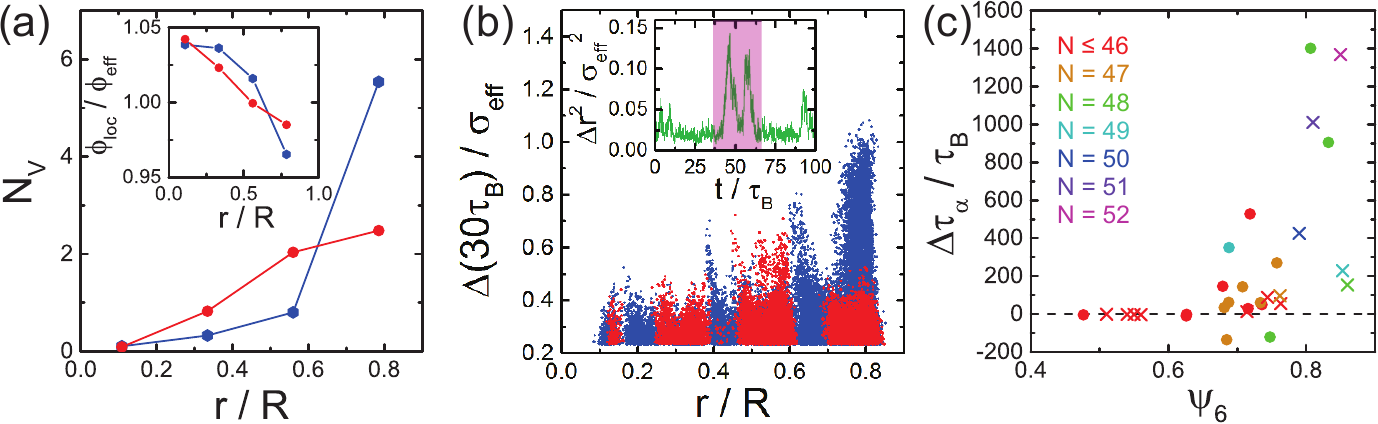}}
\caption{\label{figHexlayersdifference} (a) Average number of voids, $N_V$, as a function of radial position for experimental hexagonal (blue) and layered (red) configurations at population $N=48$. The inset shows radial local effective area fraction profile. (b) Magnitude of $10 \%$ largest displacements in time interval $30 \tau_{\mathrm{B}}$, $\Delta$, as a function of radial location for experimentally observed hexagonal (blue) and layered (red) configurations of population $N=48$. The inset shows example mean-squared displacement over $6 \tau_{\mathrm{B}}$ as measured in experiment showing large rearrangement events occuring on timescale $\sim 30 \tau_{\mathrm{B}}$ indicated by pink shaded region. (c) $\Delta \tau_\alpha = \tau_\alpha^{\mathrm{inner}} - \tau_\alpha^{\mathrm{outer}}$ as a function of temporally and spatially averaged $\psi_6$ for experimental (points) and simulated (crosses) data. Data are coloured based on corral population.}
\end{center}
\end{figure*}

Having identified novel relaxation behaviour in our model elastic pore the task remains to explain the faster-than-expected dynamics in the structurally bistable regime. A key difference between concentrically layered and hexagonal structures is the distribution of empty space within the corral. Images such as Fig. \ref{figSchematic} (c) show that enhanced hexagonal ordering leads to voids adjacent to the wall which are absent in concentrically layered systems, where the unoccupied area is distributed throughout the corral. We identify voids by considering the connectivity of empty space using an effective particle cover radius that is sufficiently large that boundary particles ``touch'' one another. A connected unoccupied region is considered a void if it has area greater than half a particle area. By considering the radial location of these voids in experiments we show that hexagonal structures indeed have an excess of voids in the layer adjacent to the boundary. This is shown in Fig. \ref{figHexlayersdifference} (a), which plots the time-averaged number of voids as a function of radial position for hexagonal (blue) and layered (red) configurations at population $N=48$. Creation of these voids adjacent to the wall thus results in a greater suppression of local area fraction in this region for hexagonal configurations compared to layered configurations as shown in the inset to Fig. \ref{figHexlayersdifference} (a). Here it is necessary to sample for short periods as over longer times $(\sim100$ $\tau_B$), the bistability means both states are sampled and any distinction is lost \cite{williams2014}.

In order to explore the dynamical differences between concentrically layered and hexagonal structures we consider particle displacements in an interval $30 \tau_\mathrm{B}$, which is observed to be an appropriate timescale for relaxation events [inset in Fig. \ref{figHexlayersdifference} (b)]. Figure \ref{figHexlayersdifference} (b) shows the magnitude, $\Delta$, of the $10 \%$ largest displacements (corresponding to $\Delta \gtrsim 0.2 \sigma_{\mathrm{eff}})$ as a function of radial position for the hexagonal (blue) and layered (red) experimental samples at population $N=48$ previously considered in (a). The hexagonal system exhibits large displacements in the layer adjacent to the boundary while, in the layered system, displacements are of a similar, smaller magnitude throughout the corral. Hexagonal configurations containing large voids near the adaptive boundary also show enhanced dynamics in this region, leading to a structural origin of the faster-than-expected relaxation shown in Fig. \ref{figEOSAngell} (b). Thus we observe two relaxation regimes which manifest as a kink in the Angell plot at the onset of structural bistability when the system begins to visit hexagonal and layered configurations.

The implication of this interpretation is that the hexagonal system exhibits spatially heterogeneous dynamics --- a fast region consisting of a single, fluid-like particle layer adjacent to the curved boundary and a slow region with enhanced hexagonal ordering deeper within the corral.  Such a decoupling of dynamics between an interfacial region and the bulk has previously been reported for confined molecular glass-formers \cite{schuller1994,pissis1998,morineau2002}, polymers \cite{krutyeva2013} and granular materials. In the latter case, hexagonal ordering has  been related to dynamically slow regions, but was enhanced at the wall quite the opposite behaviour to that found here \cite{watanabe2011}. Here, to investigate this heterogeneity, we calculate relaxation times separately for particles adjacent to the wall ($\tau_\alpha^{\mathrm{outer}}$) and particles non-adjacent to the wall ($\tau_\alpha^{\mathrm{inner}}$) and plot the difference, $\Delta \tau_\alpha = \tau_\alpha^{\mathrm{inner}} - \tau_\alpha^{\mathrm{outer}}$, as a function of time and space averaged $\psi_6$ in Fig. \ref{figHexlayersdifference} (c). Indeed, the trend in both experiment (points) and simulation (crosses) is for an increase in hexagonal ordering to result in a greater $\Delta \tau_\alpha$ characterising a stronger decoupling between dynamics in the inner and outer regions. In experiment, for a given $N$ in the bistable region, samples with higher $\psi_6$ show a greater difference between dynamics in these two regions. The non-monotonic behaviour of $\Delta \tau_\alpha$ with $N$ observed in simulation is attributed to a decrease in voidage for $N \geq 51$ as the density of the confined system is increased.

\begin{figure}
\begin{center}
\centerline{\includegraphics[width=80mm,height=63mm]{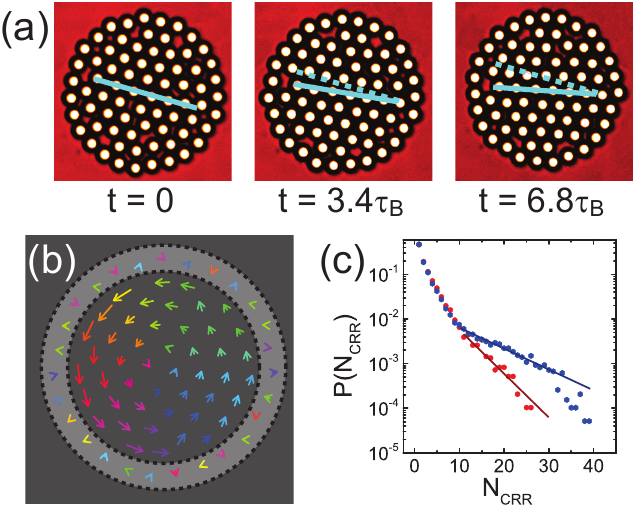}}
\caption{\label{figRearrangement} (a) Micrograph series showing co-operative rotational displacement in a hexagonal sample of population $N=47$. Cyan lines indicate the orientation of the hexagonal lattice. (b) Particle displacements in the $6.8 \tau_{\mathrm{B}}$ interval corresponding to this co-operative rotation. Arrow colour indicates direction of displacement. (c) Experimentally measured probability of observing a co-operative rearrangement of size $N_{\mathrm{CRR}}$ in locally hexagonal (blue) and concentrically layered (red) systems of population $N=48$.}
\end{center}
\end{figure}

Finally we address the co-operative mechanism by which particles relax within the corral. As shown in Figs. \ref{figRearrangement} (a) and (b), on a timescale of $7 \tau_{\mathrm{B}}$, the entire confined system undergoes a co-operative rotation about its centre. The magenta lines in Fig. \ref{figRearrangement} (a) indicate the rotation of the hexagonal configuration over the duration of the relaxation event and (b) shows the particle displacements in this interval with the colour of the arrow indicating the direction of the displacement. Large co-operative rotations of this sort occur as a result of both the voids formed in locally hexagonal structures and the nature of the confining boundary. Since the boundary is not smooth, but consists of a ring of particles, there are favoured orientations of an hexagonal configuration in which many particles adjacent to the wall sit in the gaps between a pair of wall particles. Co-operative rotations realign the hexagonal configuration from one favoured orientation to another. Thus we may picture the energy landscape as a series of 27 minima corresponding to such favoured configurations. However, it is important to note that rotations such as that pictured in Fig. \ref{figRearrangement} in which all confined particles participate are rare in experiments. More often we observe smaller co-operative rotations in which only a subsection of the system rearranges.

Relaxation is often observed to proceed via co-operative events and we propose that the formation of voids in hexagonal configurations leads to larger co-operative rearrangements. As above, we consider the largest $10 \%$ of displacements in an interval of $30 \tau_{\mathrm{B}}$ for layered and hexagonal experimental samples at population $N=48$ and identify clusters of mobile particles. Mobile particles are considered part of the same co-operatively rearranging region if their displacements begin within a cut-off distance of $1.2 \sigma_{\mathrm{eff}}$ of one another. Figure \ref{figRearrangement} (c) shows the experimentally measured probability of observing a co-operative rearrangement involving $N_{\mathrm{CRR}}$ particles in hexagonal (blue) and layered (red) configurations. For $N_{\mathrm{CRR}} \lesssim 10$ there is little difference between the two structures. However, there is an enhanced probability of observing larger co-operative rearrangements in the hexagonal system compared to the layered system. Thus the formation of voids in locally hexagonal structures leads to enhanced dynamics compared to layered systems due to a increase in the probability of observing large co-operative rearrangements. While it is true that the largest single particle displacements are observed adjacent to the wall due to the presence of large voids, the consequence of these is to enable a cascade of relaxation deeper into the confined system resulting in large co-operative rearrangements.

The observation of multiple relaxation regimes is entirely due to the deformability of the circular confining wall. Without boundary deformation, locally hexagonal configurations, as observed in the bulk supercooled liquid are forbidden. It is precisely the formation of these structures and the large voids they create near the curved boundary that allows enhanced dynamics in the wall-adjacent region resulting in large co-operative rearrangements. This result underlines the importance of the nature of the boundary when considering glassforming materials in confinement. Therefore, by facilitating the adoption of the locally preferred structure, the deformation of the wall enables the system to exhibit properties of the bulk supercooled 2d liquid.

Often when thinking about supercooled fluids one considers a rearranging subregion of a system in contact with some fixed configuration of external particles at some interface. It is through thinking along these lines that a number of theories of the glass transition were developed, including the Adam-Gibbs theory, the mosaic scenario and the random first order transition theory \cite{cavagna2009,berthier2012,adam1965,cavagna2007,kirkpatrick1989,lubchenko2007}. Furthermore, the existence of highly hexagonal subregions of comparable size to our confined system is reported in bulk two-dimensional glassformers, even with polydispersity $> 9\%$ \cite{kawasaki2007,kawasaki2011}. The structural transition in our confined system from the layered to the locally hexagonal configuration is analogous to the formation of such a medium-range crystalline subregion in contact with the external bulk liquid-like structure of the two-dimensional glassformer, modelled by our flexible, curved boundary. At the interface between the hexagonally ordered subregion and the amorphous bulk, it is natural to expect the formation of voids such as are seen adjacent to the boundary in the corral. Applying the insight gained through our model confined system then leads us to expect large co-operative rearrangements in this interfacial region leading to enhanced dynamics. Thus, although at first glance the multiple relaxation regimes reported here may seem specific to systems in curved, flexible confinement, our interpretation has wider relevance to two-dimensional glassformers in the bulk.

\section{Conclusions}

We have measured structural relaxation as a function of reduced pressure $Z$ in a model system of quasi hard discs in which two relaxation regimes are found. The confined sample is observed to relax more quickly than would be expected from an extrapolation of the lower $Z$ data or for a bulk system. Our system thus exhibits some properties of systems undergoing a fragile-to-strong transition and a liquid-liquid transition. In particular the use of reduced pressure as a parameter for supercooling reveals two relaxation regimes reminiscent of a fragile to strong transition in molecular systems  \cite{tanaka2005a,saikavoivod2005,mallamace2010}. These dynamical regimes are underpinned by the deformable boundary which results in a structural bistability at high density between concentrically layered and locally hexagonal configurations \cite{williams2013,williams2014}. Such confinement models the conceptual idea of considering the behaviour of subregions in bulk glassformers. The observed change in relaxation behaviour thus coincides with the emergence of the locally favoured structure in a 2d liquid, and we note that qualitatively similar behaviour has been observed in other bulk systems \cite{speck2012,elenius2010}.

We have identified the structural origin of this faster-than-expected dynamic behaviour. Locally hexagonal ordering leads to voids in the layer adjacent to the wall which enable large co-operative rearrangements. Our measurements open perspectives for other confined systems such as 3d hard spheres and biological systems at high packing with flexible membranes \cite{elcock2012} as well as bulk systems in which two distinct structures compete with one another. An interesting development of this research would be to tune the diameter of the confined system in order to investigate static point-to-set correlations in the supercooled fluid \cite{berthier2012}. As the lengthscale of the confinement approaches the size of co-operatively rearranging regions one expects such rearrangements to be suppressed. A further related class of systems is colloidal gels, in which particles located on the surface of the gel network branches are more mobile \cite{puertas2004, zhang2013}.

\textbf{Acknowledgements}
We thank Rob Jack for helpful discussions.
CPR and IW acknowledge the Royal Society and European Research Council (ERC Consolidator Grant NANOPRS, project number 617266). The work of ECO and HL was supported by the ERC Advanced Grant INTERCOCOS (project number 267499).

%

\end{document}